# Strong electron-boson coupling in the iron-based superconductor BaFe$_{1.9}$Pt$_{0.1}$As$_2$ revealed by infrared spectroscopy

Zhen Xing[1], Shanta Saha[2], J. Paglione[2], and M. M. Qazilbash[1,*]

[1] *Department of Physics, College of William & Mary, Williamsburg, Virginia 23187-8795, USA*

[2] *Center for Nanophysics and Advanced Materials, Department of Physics, University of Maryland, College Park, Maryland 20742, USA*

Understanding the formation of Cooper pairs in iron-based superconductors is one of the most important topics in condensed matter physics. In conventional superconductors, the electron-phonon interaction leads to the formation of Cooper pairs. In conventional strong-coupling superconductors like lead (Pb), the features due to electron-phonon interaction are evident in the infrared absorption spectra. Here we investigate the infrared absorption spectra of the iron arsenide superconductor BaFe$_{1.9}$Pt$_{0.1}$As$_2$. We find that this superconductor has fully gapped (nodeless) Fermi surfaces, and we observe the strong-coupling electron-boson interaction features in the infrared absorption spectra. Through modeling with the Eliashberg function based on Eliashberg theory, we obtain a good quantitative description of the energy gaps and the strong-coupling features. The full Eliashberg equations are solved to check the self-consistency of the electron-boson coupling spectrum, the largest energy gap, and the transition temperature ($T_c$). Our experimental data and analysis provide compelling evidence that superconductivity in BaFe$_{1.9}$Pt$_{0.1}$As$_2$ is induced by the coupling of electrons to a low-energy bosonic mode that does not originate solely from phonons.

## I. INTRODUCTION

Nearly half a century after the experimental discovery of superconductivity, Bardeen, Cooper, and Schrieffer (BCS) developed a model to explain this phenomenon [1]. The BCS mechanism provides a microscopic description of weak-coupling, phonon-mediated superconductivity in conventional superconductors. Subsequently, Eliashberg [2,3] proposed a more realistic model of the superconducting state that includes the retarded nature of the phonon induced interaction applicable to conventional strong-coupling superconductors like lead (Pb) and mercury (Hg). The agreement of the parameters in the self-consistent solutions of the Eliashberg equations, for example, in Pb, with experimental results like the phonon density of states from inelastic neutron scattering [4], electronic density of states from tunneling experiments [3], electronic heat capacity enhancement [3], and infrared absorption [5] provide strong evidence for the electron-phonon mechanism of superconductivity in conventional superconductors.

For the iron-based superconductors, it has been argued that phonons alone cannot explain the high transition temperatures [6,7]. Spin and orbital fluctuations are currently the popular candidates for mediating the formation of Cooper pairs [6,7]. There is some experimental evidence that collective spin fluctuations may be the bosons that mediate the formation of Cooper pairs. These experiments include inelastic neutron scattering studies on both electron- and hole-doped iron pnictides [8–11], scanning tunneling

mumtaz@wm.edu

spectroscopy [12], and specific heat measurements of hole (K) -doped BaFe$_2$As$_2$ [13], and quasiparticle interference imaging in LiFeAs [14].

There have been a number of infrared studies on iron-based superconductors [15–29]. However, they have not reported clear evidence of strong electron-boson coupling features in the infrared absorption data in the superconducting state normalized to the infrared absorption data in the normal state. Such features are expected to occur if superconductivity is mediated by collective bosonic excitations. Although the larger gap(s) in the iron-based superconductors are in the strong-coupling regime, only a limited number of infrared studies have considered strong-coupling approaches to model the data [17–19,25–27]. The strong-coupling methods were originally developed for strong electron-phonon interactions but they are believed to describe the coupling of electrons to any bosonic spectrum. In a few studies, researchers have obtained the electron-boson spectral density from the scattering rate only in the normal state [17,25–27]. One recent work [18] provides a method to find the electron-boson interaction both in the normal and superconducting states from the infrared scattering rate (or self-energy). However, this work does not check if the electron-boson spectral density function is self-consistent with the energy gap by solving the full Eliashberg equations. Charnukha *et al.* [19] have used a multiband Eliashberg theory to fit the optical conductivity to support the spin-fluctuation mechanism. Their model only qualitatively describes the real part of the optical conductivity in the superconducting state.

Previous experiments on high-quality single crystals of superconducting BaFe$_{1.9}$Pt$_{0.1}$As$_2$ reveal two isotropic gaps, one 2–3 meV and the other 5–7 meV [30]. Here we report infrared spectroscopy data on BaFe$_{1.9}$Pt$_{0.1}$As$_2$ that is consistent with multiband superconductivity with isotropic gaps. The important finding is that we observe strong-coupling electron-boson interaction features when the infrared absorption spectra in the superconducting state are normalized to the infrared absorption spectrum in the normal state. The frequency-dependent infrared absorption ($A$) is simply $A = 1 - R$ where the frequency-dependent infrared reflectance ($R$) is directly measured in the experiments. We identify a bosonic mode centered about 5 meV that provides the pairing glue in superconducting BaFe$_{1.9}$Pt$_{0.1}$As$_2$. We employ theoretical modeling of the absorption spectra within the Allen formalism [18,31] and Zimmermann formalism [32] based on Eliashberg theory. The full isotropic Eliashberg equations are solved to check the self-consistency of the Eliashberg function (electron-boson spectral density function), the largest energy gap, and $T_c$.

## II. SAMPLES AND EXPERIMENTS

Single crystals of BaFe$_{1.9}$Pt$_{0.1}$As$_2$ were grown using the FeAs self-flux method, which is described in Refs. [30,33] along with x-ray, transport, magnetic, and thermodynamic measurements. The dc resistivity data show the onset of superconductivity at $T_c$ = 23 K [30,33]. Magnetic susceptibility measurements show bulk superconductivity with full volume fraction [30,33]. The $ab$-plane reflectance at various temperatures from 300 to 5 K was obtained in a home-built cryogenic setup with a Bruker Vertex 80v Fourier transform infrared (FTIR) spectrometer in the frequency range 20−8000 cm$^{-1}$ (2.5−990 meV) using the technique of *in situ* gold evaporation [34]. Cryogenic ellipsometry was performed in a homebuilt vacuum chamber with a Woollam variable-angle spectroscopic ellipsometer in the energy range 0.6–6 eV [34].

## III. EXPERIMENTAL RESULTS, MODELING, AND DISCUSSION
### A. Infrared reflectance and absorption

The $ab$-plane infrared reflectance of a BaFe$_{1.9}$Pt$_{0.1}$As$_2$ crystal is shown in Fig. 1. In the normal state at $T$ = 25 K, BaFe$_{1.9}$Pt$_{0.1}$As$_2$ is highly reflective at low frequencies consistent with metallic behavior as in other metallic iron arsenides [15,20–22,24–29,34]. At $T$ = 5 K, well below $T_c$, superconductivity leads to changes in the spectrum at frequencies below ≈ 250 cm$^{-1}$. Superconductivity is observed directly from perfect reflectance at frequencies below 31.5 cm$^{-1}$ in the $T$ = 5 K spectrum. The data are consistent with a fully gapped (nodeless) superconductor close to the dirty limit [20–22,35,36]. Features at ≈260 and ≈320 cm$^{-1}$ are observed in the normal state spectrum and these features are nearly unchanged in the superconducting state spectrum. The feature at ≈260 cm$^{-1}$ is due to an infrared-active phonon. The somewhat broader feature at ≈320 cm$^{-1}$ is possibly due to a weak optical interband transition.

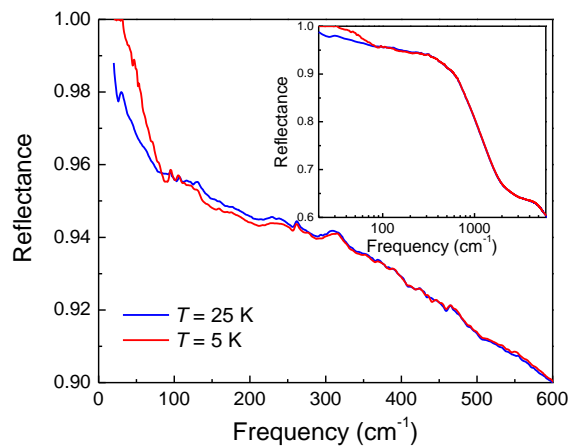

FIG. 1. The $ab$-plane infrared reflectance of BaFe$_{1.9}$Pt$_{0.1}$As$_2$ in the superconducting state ($T$ = 5 K) and normal state ($T$ = 25 K). Inset: the $ab$-plane infrared reflectance of BaFe$_{1.9}$Pt$_{0.1}$As$_2$ at $T$ = 5 K and $T$ = 25 K in a wider spectral range.

The absorption in the superconducting state $A_S(T)$ for $T < T_C$ is obtained from the equation $A_S(T) = 1 - R_S(T)$, where $R_S(T)$ is the reflectance in the superconducting state. Similarly, the normal state absorption $A_N(T = 25\,\text{K})$ is obtained from $A_N(25\,\text{K}) = 1 - R_N(25\,\text{K})$. The ratio $A_S(5\,K)/A_N(25\,K)$ is plotted as a function of frequency in Fig. 2. There are clear features at 80−200 cm$^{-1}$ which are larger than the error bars [see Fig. 2(b)]. The sharp peak at 87 cm$^{-1}$ is due to the largest gap. Above this gap feature, we observe a "valley-peak-valley" structure. When we compare our normalized infrared absorption data of BaFe$_{1.9}$Pt$_{0.1}$As$_2$ to the normalized infrared absorption data of the well-known conventional strong-coupling superconductor lead (Pb) (Refs. [5,31]), we see they are remarkably similar. In Pb, acoustic phonons are the bosonic modes which mediate the formation of Cooper pairs, and the valleys in the absorption data are due to the peaks in the phonon density of states shifted by the gap 2∆. Hence, the valleys in the absorption data of BaFe$_{1.9}$Pt$_{0.1}$As$_2$ roughly correspond to peaks in the density of states of bosonic modes shifted by the largest gap $2\Delta_3$.

In the following Secs. III B and III C, two different models have been applied to fit the normalized absorption of BaFe$_{1.9}$Pt$_{0.1}$As$_2$, in order to quantitatively determine the bosonic mode coupled to the electrons.

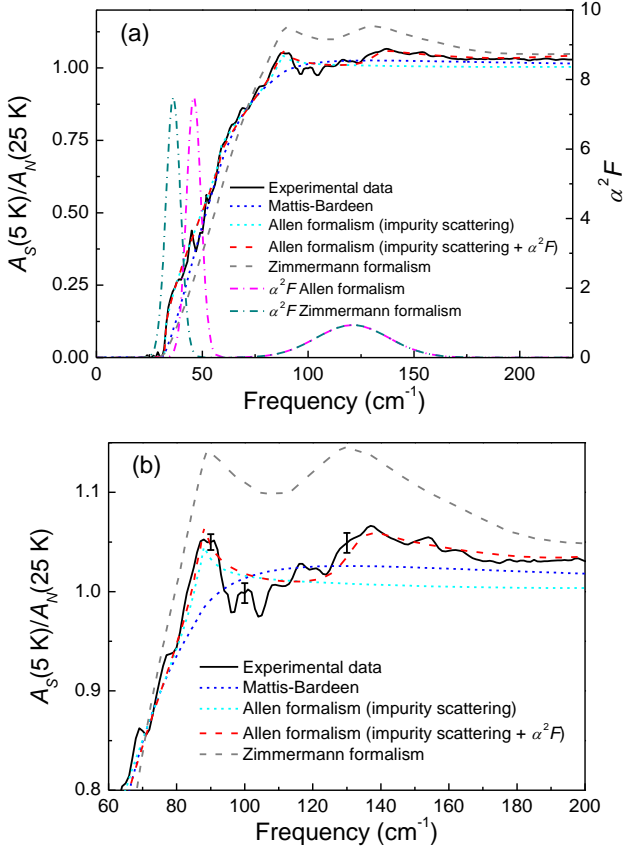

FIG. 2. (a,b) Experimental data showing infrared absorption in the superconducting state ($T$ = 5 K) normalized to infrared absorption in the normal state ($T$ = 25 K). Also shown are fits to the experimental data using four different methods described in the text. The Eliashberg functions $\alpha^2F$ shown in (a) consist of one sharp, large peak and one smaller, broad peak in the superconducting state for both Allen formalism and Zimmermann formalism. (b) Zoomed-in view of the valley-peak-valley region ($\approx$90−200 cm$^{-1}$) in the normalized absorption spectrum shown in (a). Experimental error bars at representative frequencies are also shown in (b).

## B. Modeling strong-coupling features with Allen formalism

In order to quantitatively study the bosonic modes in superconducting BaFe$_{1.9}$Pt$_{0.1}$As$_2$ and obtain a fit to the experimental normalized absorption, we start from Allen's formalism (optical self-energy method) generalized to multiband conductivity [18,31,37]. The imaginary part of the optical self-energy is

$$\Sigma_2^{op}(\omega,T) = -\frac{1}{2}\left[\int_0^\infty d\Omega\, \alpha^2F(\Omega,T) K(\omega,\Omega,T) + \frac{1}{\tau_{imp}^{op}(\omega)}\right], \quad (1)$$

where $\alpha^2F(\Omega,T)$ is the Eliashberg function (electron-boson spectral density function), $K(\omega,\Omega,T)$ is the kernel of Allen's integral equation, and $1/\tau_{imp}^{op}(\omega)$ is the impurity scattering rate [18]. Equation (1) is applicable to both the normal phase and the superconducting phase, but $K(\omega,\Omega,T)$ and $1/\tau_{imp}^{op}(\omega)$ are different for the two phases:

$$K(\omega,\Omega,T) = \frac{\pi}{\omega}\left[2\omega \coth\left(\frac{\Omega}{2T}\right) - (\omega+\Omega)\coth\left(\frac{\omega+\Omega}{2T}\right) + (\omega-\Omega)\coth\left(\frac{\omega-\Omega}{2T}\right)\right]$$

(for normal state), (2a)

$$K(\omega,\Omega,T) = \frac{2\pi}{\omega}(\omega-\Omega)\Theta(\omega-2\Delta-\Omega)$$
$$\times E\left[\frac{\sqrt{(\omega-\Omega)^2-(2\Delta)^2}}{\omega-\Omega}\right]$$

(for superconducting state at $T$ = 0 K), (2b)

where $\Theta(x)$ represents the Heaviside step function, $E(x)$ represents the complete elliptic integral of the second kind, and $\Delta$ is the energy gap. The impurity scattering rate is

$$1/\tau_{imp}^{op}(\omega) = 1/\tau_{imp} \quad \text{(for normal state)} \quad (3a)$$

$$1/\tau_{imp}^{op}(\omega) = (1/\tau_{imp})\, E\left[\frac{\sqrt{\omega^2-(2\Delta)^2}}{\omega}\right]$$

(for superconducting state at $T$ = 0 K), (3b)

in which $1/\tau_{imp}$ is a constant. Then the real part of the optical self-energy can be obtained by the Kramers-Kronig transformation:

$$\Sigma_1^{op}(\omega) = -\frac{2\omega}{\pi} P \int_0^\infty d\Omega\, \frac{\Sigma_2^{op}(\omega)}{\Omega^2-\omega^2}. \quad (4)$$

The complex optical conductivity for one channel is

$$\tilde{\sigma}(\omega) = \frac{\omega_p^2}{8\pi i}\frac{1}{\tilde{\Sigma}^{op}(\omega)-\omega/2}, \quad (5)$$

where $\omega_p$ is the plasma frequency in one channel and $\tilde{\Sigma}^{op}(\omega) = \Sigma_1^{op}(\omega) + i\Sigma_2^{op}(\omega)$. The total conductivity is the sum of different channels (here we have three channels due to the multiband nature of this material):

$$\tilde{\sigma}_{total}(\omega) = \tilde{\sigma}_{ch1}(\omega) + \tilde{\sigma}_{ch2}(\omega) + \tilde{\sigma}_{ch3}(\omega). \quad (6)$$

We then add the contributions of the interband transitions from the experimental data at higher frequencies to the total low-frequency conductivity calculated from the model. The reflectance is calculated from the real and imaginary parts of the total optical conductivity (Appendix B). The absorption is calculated from the reflectance.

In both normal state and superconducting state, the Eliashberg function $\alpha^2 F(\Omega)$ only appears in the optical self-energy of the largest gap channel, while for the two smaller gap channels only impurity scattering is considered in the optical self-energy. The parameters in the fit are as follows: the impurity scattering rate ($1/\tau_{imp} = 370$ cm$^{-1}$) consistent with the experimental data, the weights of the square of the total plasma frequency in each conductivity channel, and the three energy gaps in the superconducting state (discussed below). The total plasma frequency of 1.45 eV is obtained from the low-frequency optical conductivity data at $T = 25$ K in the normal state (Appendix A). Our best fits to the normalized absorption data and the corresponding Eliashberg function $\alpha^2 F(\Omega)$ are shown in Figs. 2(a) and 2(b). The smallest gap $2\Delta_1 = 31.5$ cm$^{-1}$ corresponds to the onset of absorption and the largest gap $2\Delta_3 = 87$ cm$^{-1}$ corresponds to the peak at 87 cm$^{-1}$ in the normalized absorption data. A third gap with energy $2\Delta_2 = 58$ cm$^{-1}$ is required to fit the shoulder around 60 cm$^{-1}$. However, $\Delta_2$ is associated with the Fermi surface with a small spectral weight (10% of the square of the normal state plasma frequency). The gaps $\Delta_1$ and $\Delta_3$ are associated with Fermi surfaces that, respectively, represent 55% and 35% of the square of the normal state plasma frequency. The smallest gap $\Delta_1$ that we observe in BaFe$_{1.9}$Pt$_{0.1}$As$_2$ is consistent with four different experiments reported in Ref. [30]. The existence of a larger gap has been previously suggested by point contact spectroscopy experiments [30]. The observation of multiple gaps is consistent with several earlier studies of other types of iron-based superconductors [19,22,38]. For an electron-doped Ba-122 system, angle-resolved photoemission spectroscopy (ARPES) data show that a small gap occurs on two electron pockets $\gamma$ and $\delta$, while a larger gap is on the outer hole pocket ($\beta$ band) [39]. The inner hole pockets are hard to observe [39,40] due to their small spectral weight. Hence $\Delta_2$ could be the gap on the inner hole pockets.

The ratio $2\Delta_3/k_B T_c = 5.44$ is clearly in the strong-coupling limit compared to the BCS weak-coupling value of 3.53. The ratios of the other two gaps to $T_c$ are either smaller than ($2\Delta_1/k_B T_c = 1.97$) or close to ($2\Delta_2/k_B T_c = 3.63$) the BCS weak-coupling value. This justifies using the Eliashberg function only in the conductivity channel associated with the largest energy gap $\Delta_3$. In order to fit the two valleys in the experimental normalized absorption spectrum, the Eliashberg function in the superconducting state consists of two Gaussian peaks: one large and sharp mode centered at frequency $\Omega_1 = 46$ cm$^{-1}$ and one broad, weaker mode centered at frequency $\Omega_2 = 121$ cm$^{-1}$. These two peaks approximately correspond to the two valleys respectively centered at frequencies 115 cm$^{-1}$ ($\approx \Omega_1 + 2\Delta_3$) and 180 cm$^{-1}$ ($\approx \Omega_2 + 2\Delta_3$) in the calculated normalized absorption spectrum. In order to obtain the correct absolute value of normalized absorption, only the weak, broad peak is necessary in the Eliashberg function for the normal state. Here we discuss the calculated normalized absorption using three methods while keeping the same energy gaps: the multi-band Allen formalism including both electron-boson interaction and impurity scattering $1/\tau_{imp}^{op}(\omega)$, the multiband Allen formalism with only impurity scattering, and multiband Mattis-Bardeen theory [41] (with constant normal state conductivity $\sigma_1 = 6000$ $\Omega^{-1}$cm$^{-1}$ consistent with the low-frequency conductivity data at $T = 25$ K shown in Appendix A). The multiband Mattis-Bardeen theory assumes the gaps are isotropic s-wave gaps in the weak-coupling limit, and the total conductivity is the superposition of the different superconducting channels. The multiband Mattis-Bardeen theory has been applied to iron-based superconductors previously [16,20–22,28,42]. The model fits are compared in Fig. 2. Neither multiple band Mattis-Bardeen theory nor the Allen formalism with only impurity scattering capture the valley-peak-valley features in the normalized absorption data. The introduction of electron-boson interaction to the optical self-energy for the largest gap is required to fit the valley-peak-valley features between $\approx$90 and 200 cm$^{-1}$.

Since the Allen formalism is expected to provide only an approximate quantitative description of strong-coupling superconductors [18,31], we take the important step to check the self-consistency of the energy gap and the Eliashberg function $\alpha^2 F(\Omega)$ used in the Allen formalism by solving the full Eliashberg equations. For this we assume an isotropic energy gap consistent with experiments [30] and the effective Coulomb pseudopotential $\mu^* = 0.1$ [43]. The Eliashberg equations are solved using EPW4.2 as described in Ref. [43]. The renormalization function $Z(\omega)$ and the superconducting gap $\Delta(\omega)$ are first solved on an imaginary energy axis and then an analytic continuation is performed to the real axis. The solved gap function is $2\Delta(\omega=0) = 85$ cm$^{-1}$, which is almost identical to the largest gap $2\Delta_3$. We also calculate $T_c$ from the Eliashberg function. The lower limit of $T_c$ can be estimated from McMillan's formula [44],

$$T_{c,\min} = \frac{\langle \omega \rangle}{1.20} \exp[-1.04(1+\lambda)/(\lambda - \mu^* - 0.62\lambda\mu^*)],$$

(7)

where $\mu^*$ is assumed to be 0.1, and

$$\lambda = 2\int_0^\infty d\Omega\ \alpha^2 F(\Omega)/\Omega, \tag{8}$$

$$\langle\omega\rangle = \{\int_0^\infty d\Omega\ \alpha^2 F(\Omega)\}/\{\int_0^\infty d\Omega\ \alpha^2 F(\Omega)/\Omega\}. \tag{9}$$

Thus we obtain $T_{c,min}$ = 17.1 K. An upper limit of $T_c$ is given by the generalized McMillan equation [18,44],

$$k_B T_{c,max} \cong 1.13\hbar\omega_{ln}\exp[-(1+\lambda)/\lambda], \tag{10}$$

where

$$\omega_{ln} = \exp[(2/\lambda)\int_0^\infty d\Omega\ \ln\Omega\ \alpha^2 F(\Omega)/\Omega], \tag{11}$$

and this gives $T_{c,max}$ = 24.6 K. The estimates of $T_c$ are consistent with the experimental transition temperature of 23 K.

## C. Modeling strong-coupling features with Zimmermann formalism

In order to confirm the results of the modeling based on the Allen formalism, we apply a second approach to model our data: the formalism of Lee, Rainer, and Zimmermann [32] (we call it Zimmermann formalism in this article) to calculate the optical conductivity in the strong-coupling regime. The Zimmermann formalism has advantages in that it is self-consistent and incorporates temperature dependence in the superconducting state. Similar results to the Zimmerman formalism have been derived by Marsiglio [45] and Schachinger and Carbotte [46], which indicate the robustness and significance of the formalism. The temperature-dependent complex conductivity in the superconducting state takes the following expression [32,47]:

$$\sigma(\omega,T) = \frac{\omega_p^2}{16\pi^3\omega}\int_{-\infty}^{+\infty}d\varepsilon\,\{\tanh\left(\frac{\varepsilon}{2k_B T}\right)M(\varepsilon,\omega)\times$$

$$[g(\varepsilon)g(\varepsilon+\omega)+h(\varepsilon)h(\varepsilon+\omega)+\pi^2] - \tanh\left(\frac{\varepsilon+\omega}{2k_B T}\right)\times$$

$$M^*(\varepsilon,\omega)[g^*(\varepsilon)g^*(\varepsilon+\omega)+h^*(\varepsilon)h^*(\varepsilon+\omega)+\pi^2]+$$

$$\left[\tanh\left(\frac{\varepsilon+\omega}{2k_B T}\right) - \tanh\left(\frac{\varepsilon}{2k_B T}\right)\right]L(\varepsilon,\omega)[g^*(\varepsilon)g(\varepsilon+\omega)+$$

$$h^*(\varepsilon)h(\varepsilon+\omega)+\pi^2]\}, \tag{12}$$

where $\omega_p$ is the plasma frequency in one conductivity channel and

$$g(\varepsilon) = \frac{-\pi\tilde{\varepsilon}(\varepsilon)}{\sqrt{\tilde{\Delta}^2(\varepsilon)-\tilde{\varepsilon}^2(\varepsilon)}}, \tag{13a}$$

$$h(\varepsilon) = \frac{-\pi\tilde{\Delta}(\varepsilon)}{\sqrt{\tilde{\Delta}^2(\varepsilon)-\tilde{\varepsilon}^2(\varepsilon)}}, \tag{13b}$$

$$M(\varepsilon,\omega) = \left[\sqrt{\tilde{\Delta}^2(\varepsilon+\omega)-\tilde{\varepsilon}^2(\varepsilon+\omega)}+\right.$$

$$\left.\sqrt{\tilde{\Delta}^2(\varepsilon)-\tilde{\varepsilon}^2(\varepsilon)}+1/\tau\right]^{-1}, \tag{14a}$$

$$L(\varepsilon,\omega) = \left[\sqrt{\tilde{\Delta}^2(\varepsilon+\omega)-\tilde{\varepsilon}^2(\varepsilon+\omega)}+\right.$$

$$\left.\sqrt{\tilde{\Delta}^{*2}(\varepsilon)-\tilde{\varepsilon}^{*2}(\varepsilon)}+1/\tau\right]^{-1}, \tag{14b}$$

in which $1/\tau$ is the impurity, scattering rate. The quantities $\tilde{\Delta}$ and $\tilde{\varepsilon}$ depend on energy $\varepsilon$, $\tilde{\varepsilon}(\varepsilon) = \varepsilon Z(\varepsilon)$ and $\tilde{\Delta} = Z(\varepsilon)\Delta(\varepsilon)$. The complex renormalization function $Z(\varepsilon)$ and superconducting gap $\Delta(\varepsilon)$ are obtained by solving the standard Eliashberg equations for isotropic systems at real energies. In eq. (12), the integral is implemented on the energy axis from negative infinity to positive infinity. The negative energy dependence of $Z(\varepsilon)$ and $\Delta(\varepsilon)$ can be obtained from the symmetry properties of $Z(\varepsilon)$ and $\Delta(\varepsilon)$. Note that the real part of both $Z(\varepsilon)$ and $\Delta(\varepsilon)$ are even functions of energy, and the imaginary parts of both $Z(\varepsilon)$ and $\Delta(\varepsilon)$ are odd functions of energy [48,49].

For the normal state, the conductivity can be expressed as

$$\sigma_N(\omega,T) = \frac{\omega_p^2}{8\pi\omega}\int_{-\infty}^{+\infty}d\varepsilon\left[\tanh\left(\frac{\varepsilon+\omega}{2k_B T}\right)-\right.$$

$$\left.\tanh\left(\frac{\varepsilon}{2k_B T}\right)\right]M_N(\varepsilon,\omega), \tag{15}$$

where

$$M_N(\varepsilon,\omega) = [-i\tilde{\varepsilon}_N(\varepsilon+\omega)+i\tilde{\varepsilon}_N^*(\varepsilon)+1/\tau]^{-1}, \tag{16}$$

and $\tilde{\varepsilon}_N(\varepsilon)$ is defined by

$$\tilde{\varepsilon}_N(\varepsilon) = \varepsilon + \int_{-\infty}^{+\infty}d\Omega\ \alpha^2 F(\Omega)\left[i\pi\coth\left(\frac{\Omega}{2k_B T}\right)-\Psi\left(\frac{1}{2}+\right.\right.$$

$$\left.\left.i\frac{-\varepsilon+\Omega}{2\pi k_B T}\right)+\Psi\left(\frac{1}{2}+i\frac{-\varepsilon-\Omega}{2\pi k_B T}\right)\right], \tag{17}$$

in which $\alpha^2 F(\Omega)$ is the Eliashberg function and $\Psi(x)$ is the digamma function. Negative energy dependence of $\alpha^2 F(\Omega)$ can also be obtained from symmetry properties of $\alpha^2 F(\Omega)$. Note that $\alpha^2 F(\Omega)$ is an odd function of frequency (energy) [50].

For the simulation based on the Zimmermann approach, the following parameters were used for the strong-coupling channel with the largest gap $\Delta_3$: $\omega_p^2$ is 35% of the square of the total plasma frequency of 1.43 eV, and the impurity scattering rate in the normal state and superconducting state is 370 and 160 cm$^{-1}$, respectively. A lower impurity scattering rate in the superconducting state compared to that in the normal state gives a better fit to the experimental data. This can be understood as follows: The effective impurity scattering rate in the superconducting state is lower because condensed electrons do not undergo impurity scattering. For weak-coupling channels with energy gaps $\Delta_1$ and $\Delta_2$, we used Mattis-Bardeen theory to calculate the conductivity. The total optical conductivity is obtained by adding up the contribution from the three parallel channels. The spectral weight (square of the plasma frequency) ratios for the three conductivity channels for the best fit are the same as in the Allen formalism (55%, 10%, and 35% for the gaps $\Delta_1$, $\Delta_2$, and $\Delta_3$). The best fit and corresponding Eliashberg function are shown in Fig. 2. It can be seen in Fig. 2 that the Zimmerman model has overall good quantitative agreement with the data because it captures the valley-peak-valley features between 90 and 200 cm$^{-1}$ and the frequencies of the peak and dip align very well with those in the experimental data. Similar to Allen's method, the Eliashberg function in the superconducting state still consists of two peaks, one large, sharp peak centered at 36.3 cm$^{-1}$ (4.5 meV), and one small, broad peak centered 121 cm$^{-1}$ (15 meV). The coupling constant $\lambda = 4.27$, and the corresponding upper limit transition temperature $T_c$ is 20.5 K. Analogous with the results of the Allen formalism, only the small, broad peak is included in the Eliashberg function for calculating the normal state conductivity. The result of solving the Eliashberg equations at 5 K gives the gap function $2\Delta(\omega=0) = 81.2$ cm$^{-1}$, which is close to the result using Allen's formalism.

Our models based on the Allen and Zimmermann formalisms quantitatively describe the energy gaps and the strong-coupling features in the experimental data (see Fig. 2). However, we note that the model based on the Allen formalism gives a better fit to the experimental data compared to the model based on the Zimmermann formalism.

### D. Origin of the bosonic modes

Next we discuss the origin of the two peaks in the Eliashberg function. The promising candidates for bosons which mediate the formation of Cooper pairs are either spin fluctuations or orbital fluctuations (induced by Fe phonons). Spin resonance modes have been determined by inelastic neutron scattering experiments [8–11]. The spin resonance, which is observed only in the superconducting state in cuprates, heavy-fermion, and iron-based superconductors, is generally considered a feedback effect from superconductivity. Despite some theoretical controversies, the resonance is viewed as a spin-exciton bound state in the particle-hole channel. The appearance of the resonance implies a sign change of superconducting gap(s) between either different patches of the Fermi surface or different Fermi pockets connected by a resonance mode at momentum $q$ (see Ref. [51] and references therein). From the modeling of our infrared absorption data, the large sharp peak in the Eliashberg function of BaFe$_{1.9}$Pt$_{0.1}$As$_2$ is centered at $5.1 \pm 0.6$ meV ($41 \pm 5$ cm$^{-1}$), with a full width at half maximum of 1 meV, and is only present in the superconducting state. We note that the spin resonance mode at 3D antiferromagnetic ordering wave vector $Q = (1, 0, -1)$ occurs in BaFe$_{1.9}$Ni$_{0.1}$As$_2$ (a superconductor with $T_c = 20$ K and similar to BaFe$_{1.9}$Pt$_{0.1}$As$_2$), with resonance energy $\hbar\omega_{res} = 7 \pm 0.5$ meV, and width $d = 1.9 \pm 0.7$ meV [8]. Inelastic neutron scattering experiments on BaFe$_{1.9}$Pt$_{0.1}$As$_2$ are not available at present. If the bosonic mode we have observed is due to spin fluctuations, then we expect that a spin resonance mode about 5 meV will be observed in future inelastic neutron scattering experiments. The center frequency of the bosonic mode in our infrared experiments is also not that different from the spin resonance mode of another electron-doped material Ba(Fe$_{1-x}$Co$_x$)$_2$As$_2$ which is $\sim$8–9 meV [10,11]. Note that the bosonic mode observed in the optical response is the $q$ averaged (all momenta in the Brillouin zone) local susceptibility. From the above discussion, we infer that the sharp peak about 5 meV in the Eliashberg function of BaFe$_{1.9}$Pt$_{0.1}$As$_2$ possibly represents the spin resonance in the superconducting state. The important point is that the 5-meV peak cannot be due to phonons alone because it is lower in energy compared to the energy of the lowest peak in the phonon density of states in the parent compound or doped BaFe$_2$As$_2$ [52,53]. Moreover, since phonons are present in both the normal and superconducting states, the 5-meV peak cannot be due to phonons alone because it is only required in modeling the superconducting state data and not required for modeling the normal state data.

The broad, weak peak in $\alpha^2F(\Omega)$ is centered at 15 meV (121 cm$^{-1}$), with a width of 5 meV, and is required in the models for both the superconducting and normal states. Inelastic x-ray scattering experiments have measured the lowest-energy peak in the Fe phonon density of states centered at 13 meV, with width approximately 5 meV. The phonon density of states is nearly temperature independent [54]. Phonons are likely the origin of the weak, broad mode. Actually, the position and the width of the broad peak are also very similar to the prediction of the resonance peak of the s$_{++}$ wave pairing state [55]. Possible explanations are that the weak, broad mode is either due to electron-phonon interaction or due to phonon induced orbital fluctuations. Note that the total electron-boson coupling constant $\lambda = 3.5 - 4.3$ contains a significant contribution of $2.8 - 3.6$ from the sharp peak, and a minor contribution of only 0.7 from the broad peak. If the sharp peak in the Eliashberg function is due to spin fluctuations, this means spin fluctuations play the dominant role in superconductivity in $BaFe_{1.9}Pt_{0.1}As_2$. It would also support the presence of a predominant $s_\pm$ gap in superconducting $BaFe_{1.9}Pt_{0.1}As_2$ [6]. However, we note that superconductivity with relatively high $T_c$ is preserved in the presence of large impurity scattering in $BaFe_{1.9}Pt_{0.1}As_2$. This is more consistent with an $s_{++}$ pairing state because the $s_\pm$ pairing state is expected to be fragile against impurities due to interband scattering [56].

### E. Temperature dependent features

Finally, we study the temperature dependence of the normalized absorption spectra. The absorption spectra in the superconducting state at $T$ = 5, 10, 15, and 20 K, are normalized to the normal state absorption data ($T$ = 25 K) and plotted in Fig. 3(a). It is clear that the amplitude of the strong-coupling features due to electron-boson interaction decreases when temperature increases toward $T_c$. However, there is little frequency dependence of these features for temperatures at and below 15 K. At $T$ = 20 K, still below $T_c$, the strong-coupling features weaken further and move to lower frequencies. This may be caused by a reduction of the energy gap $\Delta_3$ magnitude and a downward shift in center frequency $\Omega_1$ of the bosonic peak as the temperature approaches $T_c$ from below. The Allen formalism for the superconducting state is meant for $T$ = 0 K and works well for temperatures much below $T_c$. To the best of our knowledge, the Allen formalism for the superconducting state at higher temperatures does not exist at present. Hence, we cannot quantitatively model the temperature dependence of the bosonic mode based on the Allen formalism. Nevertheless, we attempt to follow the temperature dependence of the energy gaps using two alternative methods discussed below. The first method is based on Mattis-Bardeen theory. The second method based on the Zimmerman formalism also allows us to model the temperature dependence of the low-energy bosonic mode.

In the first method, the temperature-dependent energy gap $2\Delta_3(T)$ is estimated directly from the normalized absorption because it corresponds to the first prominent peak position [shown by arrows in Fig. 3(b)] and is plotted in Fig. 3(c). The temperature dependence of $\Delta_1$ and $\Delta_2$ cannot be obtained directly from the data. However, since the ratio $2\Delta/k_BT_c$ for the smaller two gaps shows they are in the weak-coupling regime, we have modeled the normalized absorption using three-band Mattis-Bardeen formalism (we assume the temperature dependence of the largest gap can be modeled with Mattis-Bardeen theory). The results are shown in Fig. 3(c) with hollow symbols. The largest and smallest gaps appear to deviate from the BCS prediction close to $T_c$.

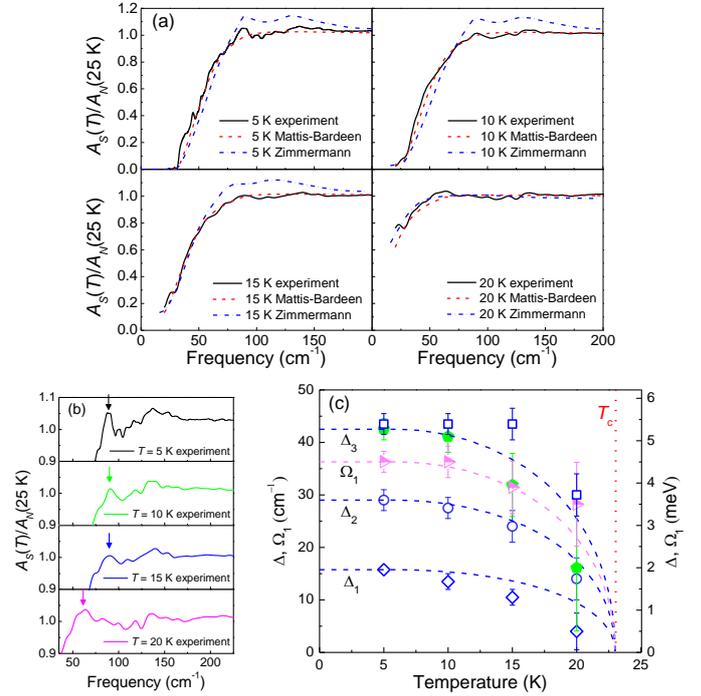

FIG. 3. (a) Solid lines are temperature-dependent infrared absorption in the superconducting state normalized to infrared absorption in the normal state at $T$ = 25 K. Dashed lines (red) are Mattis-Bardeen fits to the normalized infrared absorption data. Dash-dotted lines (blue) are the fits using Zimmermann's formalism for the largest energy gap, and Mattis-Bardeen formalism for the two smaller energy gaps. (b) Zoomed-in view of the spectra showing the peak associated with the largest gap $2\Delta_3$ and the valley-peak-valley strong-coupling features at different temperatures in the superconducting state. Arrows indicate the frequency of the first prominent peak in the normalized absorption spectrum due to the energy gap $2\Delta_3$ in the presence of impurity scattering. (c) Plot of the temperature dependence of the three energy gaps and bosonic mode $\Omega_1$. Hollow symbols (blue) represent energy gaps from Mattis-Bardeen formalism (see text), filled symbols (green) represent the energy gap $\Delta_3$ from Zimmermann formalism, and half-hollow symbols (magenta) represent bosonic mode $\Omega_1$. The dashed lines are the BCS prediction of the temperature dependence of the energy gaps. The vertical dotted line represents $T_c$.

Since the Mattis-Bardeen description does not capture the temperature dependence of the strong-coupling features and the low-energy bosonic mode, we attempt to fit the temperature-dependent normalized absorption using Zimmermann's formalism for the largest gap channel. In the modeling, we assume the low-energy bosonic mode is temperature dependent and follows a similar functional dependence as the energy gap [10]. Temperature-dependent complex renormalization function $Z(\varepsilon)$ and superconducting gap $\Delta(\varepsilon)$ are obtained by solving the standard Eliashberg equations for isotropic systems at real energies. The Zimmermann formalism is applied in the largest energy gap channel, and the temperature dependence of the two smaller gaps in the weak-coupling regime are modeled using Mattis-Bardeen theory. The simulation results are shown in Fig. 3(a). The theoretical model roughly captures the temperature-dependent trend of the valley-peak-valley features. At $T$ = 10 and 15 K, the valley-peak-valley features become weaker compared to $T$ = 5 K simulation, while there is some frequency dependence at $T$ = 15 K compared to the $T$ = 5 and 10 K simulations. At $T$ = 20 K, a temperature close to $T_c$, the valley-peak-valley features are nearly washed out in the simulation consistent with the experimental data. The temperature dependence of the energy gaps and the bosonic mode from the model is shown in Fig. 3(c). There are larger error bars at higher temperatures due to uncertainty in the solution of the Eliashberg equations using the EPW software when the temperature approaches $T_c$.

## IV. CONCLUSION

To conclude, we have observed temperature-dependent features in the infrared absorption spectra arising from the energy gaps and strong electron-boson interaction in the superconductor BaFe$_{1.9}$Pt$_{0.1}$As$_2$. This was enabled by careful, systematic cryogenic infrared reflectance measurements. The data are consistent with three nodeless energy gaps in the superconducting state, out of which only the largest gap is in the strong-coupling regime. We obtain the Eliashberg function (electron-boson spectral density function) by modeling the absorption data with both the generalized Allen formalism and Zimmermann formalism. The largest gap, the $T_c$, and the Eliashberg function were verified to be self-consistent within the Eliashberg theory. We find that superconductivity in BaFe$_{1.9}$Pt$_{0.1}$As$_2$ arises primarily due to pairing of electrons induced by a bosonic mode centered at 5.1 ± 0.6 meV. This bosonic mode cannot be attributed to phonons alone because it occurs at an energy less than the lowest-energy peak in the phonon density of states. The bosonic mode may originate from spin fluctuations although we cannot rule out the role of orbital fluctuations or another mechanism.

## Acknowledgments

M.M.Q. acknowledges support from the National Science Foundation (Award No.# NSF DMR-1255156) and the Virginia Space Grant Consortium for the research conducted at William & Mary. Research at the University of Maryland was supported by the Air Force Office of Scientific Research (Award No. FA9550-14-1-0332) and the National Science Foundation Division of Materials Research (Award No. DMR-1610349). Z.X. is grateful to Hao Shi for his assistance with running the EPW4.2 program in Linux environment.

## APPENDIX A: OPTICAL CONDUCTIVITY

The temperature dependence of the real part of the optical conductivity $\sigma_1$ is shown in Fig. 4. It is obtained from Kramers-Kronig transformation of the reflectance data constrained by cryogenic ellipsometry data, similar to the procedure described in Ref. [34]. At $T$ = 5 K, the real part of the conductivity is negligible below the frequency 31.5 cm$^{-1}$, corresponding to the smallest gap. At higher frequencies, there is a sharp increase of the conductivity just above the gap and subsequently the conductivity reaches a maximum, which is a clear indication of superconductivity in the dirty limit. Indeed, the scattering rate in the normal state ($T$ = 25 K) is 370 cm$^{-1}$ which is much larger than the energy gaps indicating that superconductivity is in the dirty limit. In fact, the large radius Pt ion doped into the FeAs$_4$ tetragon leads to significant impurity scattering and to some degree of localization at higher temperatures in the normal state. This can be seen from the nonmonotonic frequency dependence of $\sigma_1$ at low frequencies in the normal state at higher temperatures (see Fig. 4).

The inset in Fig. 4 clearly shows the "missing" spectral weight between the normal state conductivity and the superconducting state conductivity. The missing spectral weight in the superconducting state is transferred into the delta function at zero frequency which represents the superfluid response to a dc electric field. The missing spectral weight area is equal to the superfluid density [22],

$$\omega_{ps}^2 = 8 \int_0^{\omega_c} d\omega [\sigma_1(\omega, T = 25K) - \sigma_1(\omega, T = 5K)] =$$

$1.9 \times 10^7$ cm$^{-2}$, where the cutoff frequency $\omega_c$ = 400 cm$^{-1}$ is chosen so that the integral converges smoothly. The superfluid density is consistent with that obtained from the low-frequency limit $\omega_{ps}^2 = -\omega^2 \varepsilon_1 (\omega \to 0)$, where $\varepsilon_1$ is the real part of the dielectric function [22,57]. We use the Drude-Lorentz model to separate the contribution of free carriers and interband transitions to the conductivity in the normal state ($T = 25$ K) [57]. In the simplest Drude-Lorentz model, a single Drude feature is sufficient to describe the free carrier contribution. The superfluid density at $T = 5$ K is 14% of the Drude spectral weight in the normal state ($T = 25$ K). An interpretation is that 14% of free carriers in the normal state have condensed into the superconducting state.

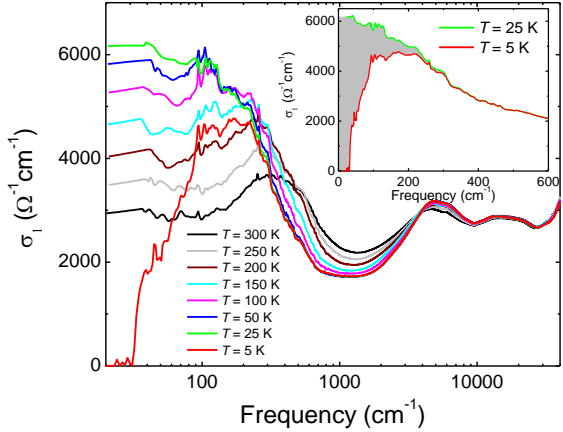

FIG. 4. The real part of the $ab$-plane optical conductivity $\sigma_1$ is plotted as a function of frequency at different temperatures. Inset: the region shaded gray is the "missing area" between the normal and superconducting state real conductivity that moves into the delta function at $\omega = 0$ in the superconducting state.

## APPENDIX B: ABSOLUTE REFLECTANCE AND ABSORPTION CALCULATED USING THE TWO MODELS

Absolute reflectance and absorption calculated [57] from the total optical conductivity based on the Allen formalism and the Zimmermann formalism in the superconducting state ($T = 5$ K) and normal state ($T = 25$ K) are shown in Fig. 5. We have obtained quantitatively good agreement to the absolute reflectance and absorption data using the Allen formalism. The Zimmermann formalism agrees better with the experimental data at lower frequencies compared to higher frequencies (above ≈100 cm$^{-1}$).

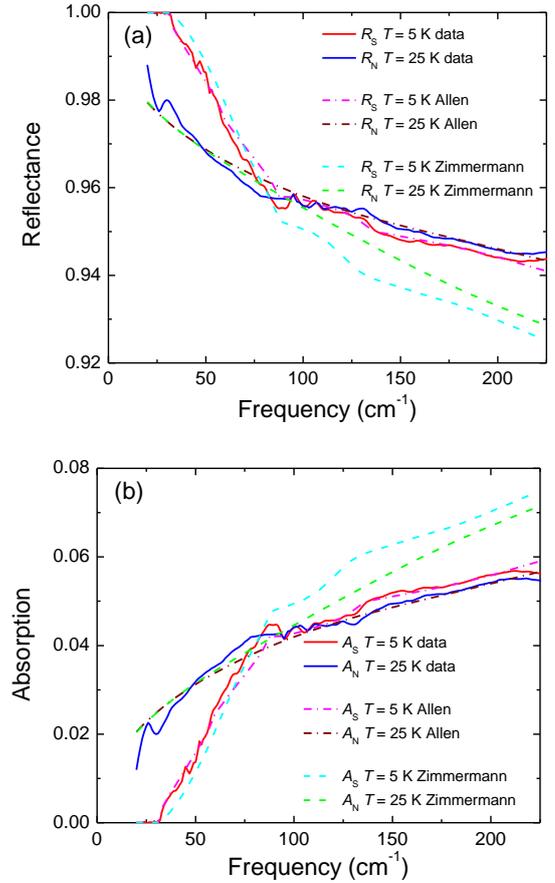

FIG. 5. The frequency-dependent (a) reflectance and (b) absorption in the superconducting state (5 K) and the normal state (25 K) calculated from the Allen formalism and the Zimmermann formalism and compared to the experimental data.